\documentclass[conference]{IEEEtran}
\IEEEoverridecommandlockouts
\usepackage{cite}
\usepackage{amsmath,amssymb,amsfonts}
\usepackage{algorithmic}
\usepackage{graphicx}
\usepackage{textcomp}
\usepackage{url}
\usepackage{enumerate}
\usepackage[ruled,linesnumbered]{algorithm2e}
\usepackage{algorithmic}
\usepackage{subfigure}
\usepackage{float}
\usepackage{amssymb}
\usepackage{graphicx}
\usepackage{epstopdf}
\def\BibTeX{{\rm B\kern-.05em{\sc i\kern-.025em b}\kern-.08em
    T\kern-.1667em\lower.7ex\hbox{E}\kern-.125emX}}
\begin{document}

\title{Learning Deterministic Policy with Target \\for Power Control in Wireless Networks
}
\author{\IEEEauthorblockN{Yujiao Lu$^1$, Hancheng Lu$^1$, Liangliang Cao$^2$, Feng Wu$^1$, Daren Zhu$^1$}
\IEEEauthorblockA{$^{1}$Department of Electrical Engineering and Information Science, \\
University of Science and Technology of China, Hefei, Anhui 230027 China \\
$^{2}$Columbia University and Yahoo Labs, New York, the United States\\
lyj66@mail.ustc.edu.cn, hclu@ustc.edu.cn, liangliang.cao@gmail.com, fengwu@ustc.edu.cn, darenzhu@mail.ustc.edu.cn}
}

\maketitle

\begin{abstract}
\emph{Inter-Cell Interference Coordination} (ICIC) is a promising way to improve energy efficiency in wireless networks, especially where small base stations are densely deployed. However, traditional optimization based ICIC schemes suffer from severe performance degradation with complex interference pattern. To address this issue, we propose a \emph{Deep Reinforcement Learning with Deterministic Policy and Target} (DRL-DPT) framework for ICIC in wireless networks. DRL-DPT overcomes the main obstacles in applying reinforcement learning and deep learning in wireless networks, i.e. continuous state space, continuous action space and convergence. Firstly, a \emph{Deep Neural Network} (DNN) is involved as the actor to obtain deterministic power control actions in continuous space. Then, to guarantee the convergence, an online training process is presented, which makes use of a dedicated reward function as the target rule and a policy gradient descent algorithm to adjust DNN weights. Experimental results show that the proposed DRL-DPT framework consistently outperforms existing schemes in terms of energy efficiency and throughput under different wireless interference scenarios. More specifically, it improves up to 15\% of energy efficiency with faster convergence rate.
\end{abstract}


\section{Introduction}
\emph{Reinforcement learning} (RL) is provided with the ability of learning a rewarding behavior in a pre-unknown environment via trial-and-error \cite{b3}. It has been successfully applied in many physical tasks. Recently, significant advances in RL have been obtained by combining \emph{deep learning} (DL) into RL, resulting in \emph{deep reinforcement learning} (DRL), mainly represented by the algorithms of ``Deep Q Network'' (DQN) \cite{b4} and ``Deep Deterministic Policy Gradient'' (DDPG) \cite{b5}. DRL is capable of human-level performance by using deep neural network function approximators.

In modern wireless networks, especially where small cell networks are densely deployed, the interferences between co-frequency cells are getting heavier and heavier, making \emph{inter-cell interference coordination} (ICIC) essential to improve the system performance. While traditional optimization-based ICIC schemes are hand-crafted and suffer degradation with complex interference patterns \cite{b20}, RL is promising to achieve the breakthrough \cite{b12}. However, there are several challenges to apply RL in wireless networks. First, the state and action spaces in wireless networks are uncountable, making the tabular-represented Q learning  \cite{b13} and the action-discrete DQN infeasible. Both of them base on the iteration of Q-value, which limits the problem space. The observations of wireless networks range in a large scale, varying with the distribution of users and the dynamic power allocation. And the action space, which refers to power allocations in ICIC, is continuous because of the finer-grained requirements \cite{b14}. Although DDPG has been proposed to solve the continuous control problems, it suffers with the slow convergence due to the neural networks approximated actor-critic model. All of them are value-based, which means they are impossible to learn fast since they have to update the values by large amounts of iterations. Also they may suffer from oscillations when converge. Another interesting thing is that a lot of excellent analyses about wireless networks have been done and they may provide guidance to the learning methods as something like ``teacher's advice''. While the existed RL algorithms cannot leverage the well-known prior knowledge obviously.
\begin{figure}[h]
\centering
\includegraphics[width=7cm, height=8cm]{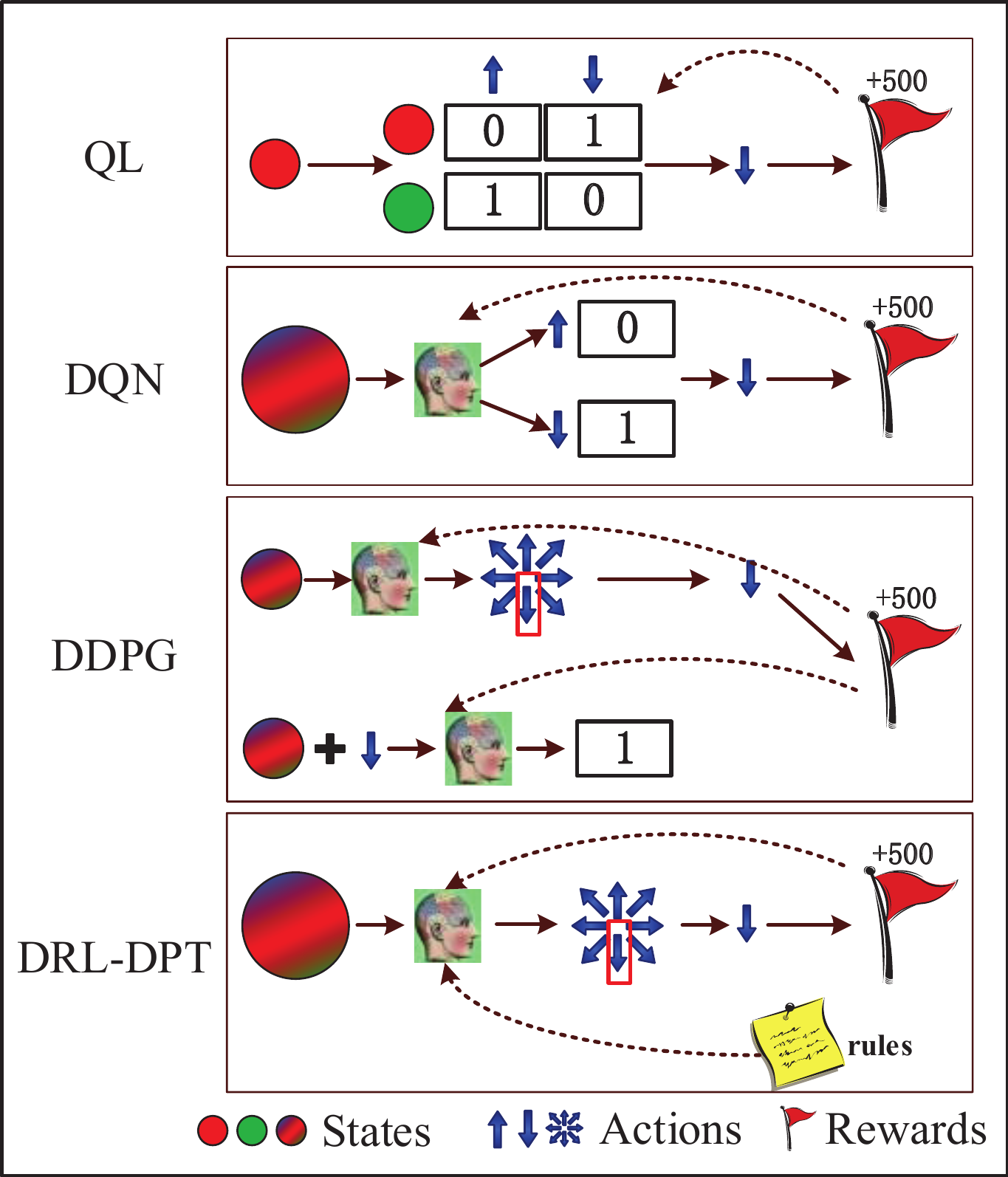}
\caption{Comparing the differences among QL, DQN, DDPG and DRL-DPT. Ordinary Q learning is a value-based and table-presented algorithm which solves simple problems with discrete and low-dimensional spaces. It learns by updating a Q-table and selects the action making the Q-value maximum at each iteration. DQN adapts the internal of QL to continuous high-dimensional state space by replacing the action-value table with a deep neural network. It observes the environment directly without knowing all the possible states in advance. At each iteration, it computes Q-values for every possible actions and choose the maximum one. However, DQN still can't deal with high-dimensional action space. DDPG extends the idea of neural network function approximators in DQN and develops the ability for continuous control. It uses two neural networks to consist the actor-critic model. The actor network generates an action according to its observation and the critic generates a Q-value for this state-action tuple. The proposed DRL-DPT is a policy-based algorithm using a deep neural network as actor to obtain continuous control. It is notable that DRL-DPT not only learn through autonomous exploration but also guidance given by a target rule.}
\label{compare}
\end{figure}

In this paper, we propose a novel \emph{deep reinforcement learning with deterministic policy and target} (DRL-DPT) framework for ICIC in wireless networks. It is shown that the energy-efficiency of different wireless networks improves greatly by using DRL-DPT. In fact, DRL-DPT could obtain up to 15\% increase in terms of the energy-efficiency and the system throughput compared to baselines. DRL-DPT uses a \emph{deep neural network} (DNN) to play the role of actor underlying the success of DDPG to deal with the continuous state and action spaces. DRL-DPT is policy-based and do not rely on the update of Q-values. It gives deterministic power allocation schemes at each iteration without any extra computation. It also adapts the mature theories in wireless networks to help build the optimization rules, aiming at higher energy-efficiency \cite{b6}. A dedicated reward function is designed to guide the learning process based on the \emph{deterministic policy gradient} (DPG) algorithm \cite{b7}. By removing the critic network and learning directly from the expected objective, DRL-DPT guarantees the convergence with an acceptable speed, as well as good stability. A simple compare between Q learning, DQN, DDPG and DRL-DPT can be found in fig.~\ref{compare}.

The rest of this paper is organized as follows. Section 2 briefly introduces some backgrounds on RL and DRL. Section 3 introduces the considering application scenario and some useful acknowledgements. In section 4, illustrations about the proposed DRL-DPT are given in detail. Section 5 describes the experimental results. Finally in section 6, we conclude our work with discussion about future work.

\section{Background}

RL is a process where an agent interacts with its environment, receiving observations and selecting actions to maximize a reward signal provided by the environment. It has made many achievements in robotic tasks. Recently, some literatures also adapt RL to communication systems.

RL algorithms typically leverage the \emph{Markov decision process} (MDP) formulation. In an MDP, $\mathbb{A}$ is a set of actions and $\mathbb{S}$ is a set of states. There are two functions within this process: a transition function $(T\!: \mathbb{S}\!\times\! \mathbb{A} \!\mapsto\! \mathbb{S})$ and a reward function $(r\!: \mathbb{S} \!\times\! \mathbb{A} \!\mapsto\! r)$. The return from a state is often described as the sum of discounted future rewards: $R_t\!=\!\sum_{i=t}^T \gamma^{i-t}r_t(s_i,a_i)$, where $\gamma \!\in\! [0,1]$ is a discount factor.

The action-value function is used in many RL algorithms to describe the expected return after taking an action $a_t$ in state $s_t$ following policy $\pi$:
\begin{equation}
Q^{\pi}(s_t,a_t) = \mathbb{E}[R_t|s_t,a_t]
\end{equation}

Many approaches in RL make use of the recursive relationship known as the Bellman equation:
\begin{equation}
Q^{\pi}(s_t,a_t) = \mathbb{E}[r(s_t,a_t)+\gamma \mathbb{E} Q^{\pi}(s_{t+1},a_{t+1})]
\end{equation}

It is possible to learn the optimal policy $\pi^*$ using transitions from a different stochastic behavior policy.
Different RL algorithms have different ways to learn the optimal policy which maximizing the reward. \emph{Q Learning}, a widely used off-policy algorithm, uses the greedy policy: $\pi^*(s) = argmax_a Q(s,a)$. In DQN and DDPG, function approximators parameterized by $\theta$ are used, and they optimize by minimizing the loss:
\begin{equation}
L(\theta) = \mathbb{E}[(Q(s_t,a_t|\theta)-y_t)^2]
\end{equation}
where $y_t = r(s_t,a_t) + \gamma Q(s_{t+1},\pi^*(s_{t+1})|\theta)$ is calculated by another target network.

In this paper, we denote the discount factor to be 0 since the considered ICIC problem is some how instant. We approximate the policy by a neural network, too. But when optimizing, we directly aim at maximizing the feedback reward $R_t$ by formulating the reward function theoretically. In practice, the optimization is done by minimizing the cost $C(\theta)$:
\begin{equation}
C(\theta) \varpropto (1/R_t)
\end{equation}

\section{System Model}

\begin{figure}[t]
\centering
\includegraphics[width=7cm, height=5.5cm]{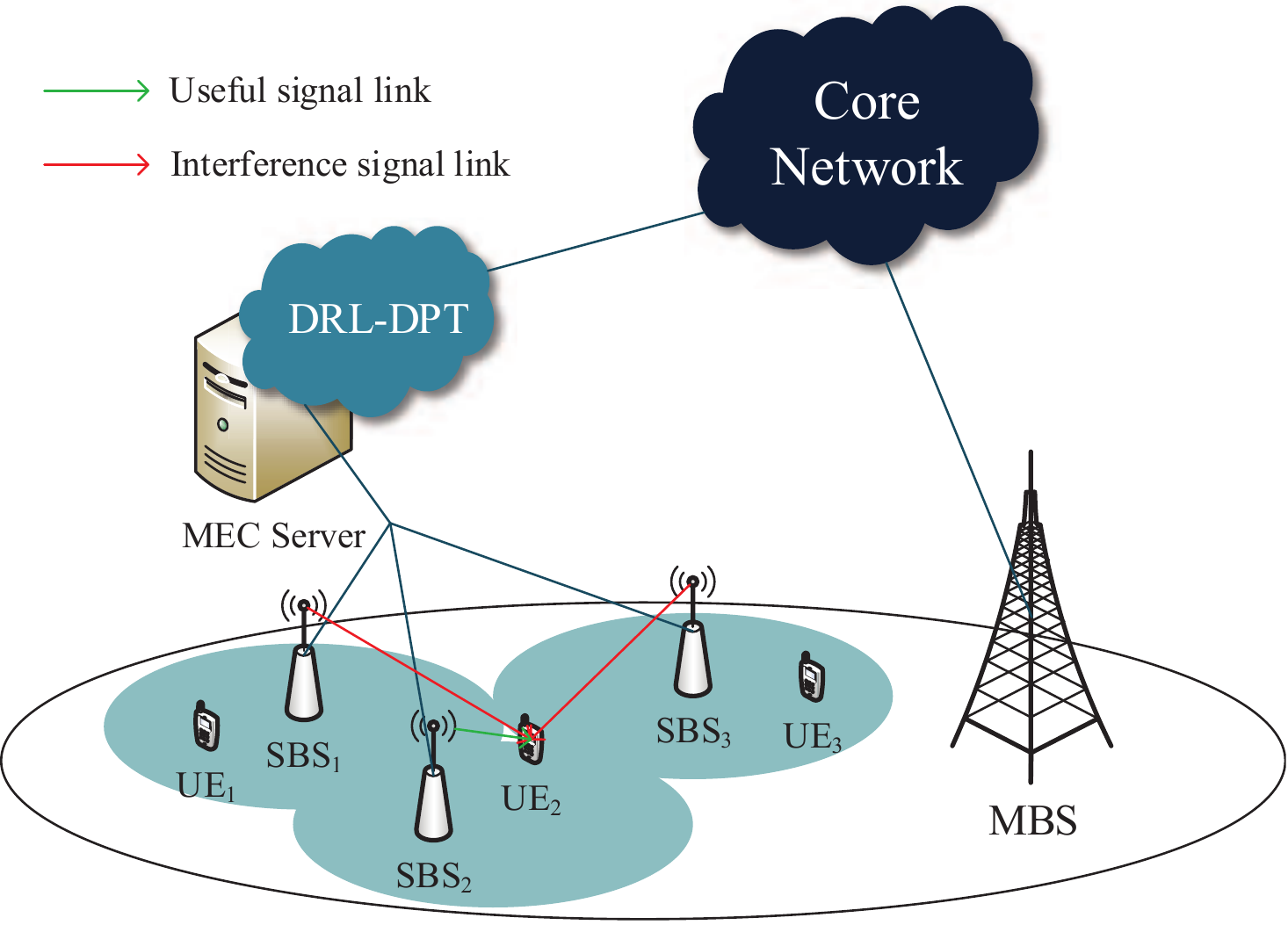}
\caption{The considering application scenario. A two tier heterogeneous network is considered and the inter-cell interferences between small base stations are focused on.}
\label{scenario}
\end{figure}

In order to describe the proposed framework clearly, we demonstrate the scenario and assumptions briefly in this section.
As shown in Fig.~\ref{scenario}, modern communication systems often overlay the small base stations (SBSs) based on the orthogonal macrocell frequency to enlarge the system capacity, however, such practices will bring heavy inter-cell interferences \cite{b16}.
To handle such a challenge, this paper proposes DRL-DPT algorithm to optimize inter-cell interference coordination by doing. Our algorithm runs on {\itshape mobile edge computing} (MEC) server \cite{b17}.

To depict the inter-cell interference mathematically, we assume $M$ SBSs and $N_m$ users connected to SBS $m$.
At time slot $l$, user $n$ gets interested signal from its associated SBS $m$, the signal strength can be described as
\begin{equation}
S_{m,n,l} = P_{m,l} \cdot g_{m,n,l}
\label{eq1}
\end{equation}
where $P_{m,l}$ represents the transmission power of SBS $m$ at time slot $l$, $g_{m,n,l}$ represents the channel gain from SBS $m$ to user $n$.

Similarly, user $n$ receives interference signals $S_{k,n,l} (k \!\not=\! m)$ from other SBSs.
According to Shannon theorem, the {\itshape signal to interference and noise ratio} (SINR) and downlink data rate of user $n$ can be expressed as
\begin{equation}
SINR_{m,n,l} = \frac{P_{m,l}g_{m,n,l}}{\sigma_{m,n,l}^2+\sum_{1 \leq k \leq M, k \not= m} P_{k,l}g_{k,n,l}}
\label{eq2}
\end{equation}
\begin{equation}
R_{m,n} = \frac{1}{LT_{sf}} \sum_{l=1}^L T_{sf}Wlog(1+SINR_{m,n,l})
\label{eq3}
\end{equation}
where $\sigma_{m,n,t}^2$ is the noise. $L$ represents the length of a frame and $T_{sf}$ is the duration of a subframe. $W$ represents the bandwidth assigned to user $n$.

The system power consumption is evaluated by the power model introduced in \cite{b6}, which is expressed as
\begin{equation}
E_{m,l} = \bigtriangleup pP_{m,l} + p_0 , 0 \leq P_{m,l} \leq P
\label{eq4}
\end{equation}
where $\bigtriangleup p$ is a parameter related to power amplifier. $p_0$ represents the circuit power and $P$ stands for the maximum power the SBSs could choose.
The system performance is graded by energy efficiency, which is defined as the ratio of system throughput and system power consumption \cite{b18}:
\begin{equation}
\eta = \frac {\sum_{m=1}^M \sum_{n=1}^{N_m} R_{m,n}} {\sum_{m=1}^M \sum_{l=1}^L E_{m,l}}
\label{eq5}
\end{equation}

To implement ICIC and enhance the energy efficiency, the DRL-DPT algorithm does power control to schedule the interference signals. It observes the system and gives specific power allocation for every SBS at each frame. Detailed process can be found in the next section.

\section{Proposed Algorithm}

In this section, DRL-DPT is illustrated in detail. Firstly, we formulate the optimization objective based on the widely used theories in wireless networks. It provides DRL-DPT with a learning target. Then in the next subsection, we present and explain the structure of DRL-DPT. Finally, we demonstrate the learning mechanism guided by the target rule and account for the no-label gradient descent executed on the actor.

\subsection{Target}

In this paper, we focus on maximizing the system energy efficiency to trade off throughput and power consumption.
The optimization problem can be formulated as
%\begin{scriptsize}
\begin{subequations}
\begin{align}
\max ~~~ &\eta \\
\text s.t. ~~~~~ &R_{m,n} \ge R_n^{th}, \forall m \in \mathbb{M}, \forall n \in \mathbb{N}_m\\
            &0 \le P_{m,l} \le P, \forall m \in \mathbb{M}, \forall l \in \mathbb{L}
\end{align}
\label{eq6}
\end{subequations}
%\end{scriptsize}
where $R_n^{th}$ refers to the throughput requirement of user $n$.
Respectively, $\mathbb{M}\!=\!\{1,2,...,M\}, \mathbb{N}_m\!=\!\{1,2,...,N_m\}, \mathbb{L}\!=\!\{1,2,...,L\} $.
(\ref{eq6}b) indicates the demand that users' requirements have to be satisfied.
(\ref{eq6}c) means the transmission powers are positive numbers less than the predefined upper limit.
When designing the modules of DRL-DPT, we consider the maximization objective as well as two constraints.

\subsection{Neural Network Approximated Actor}

\begin{figure*}[htb]
\centering
\includegraphics[width=16cm, height=6cm]{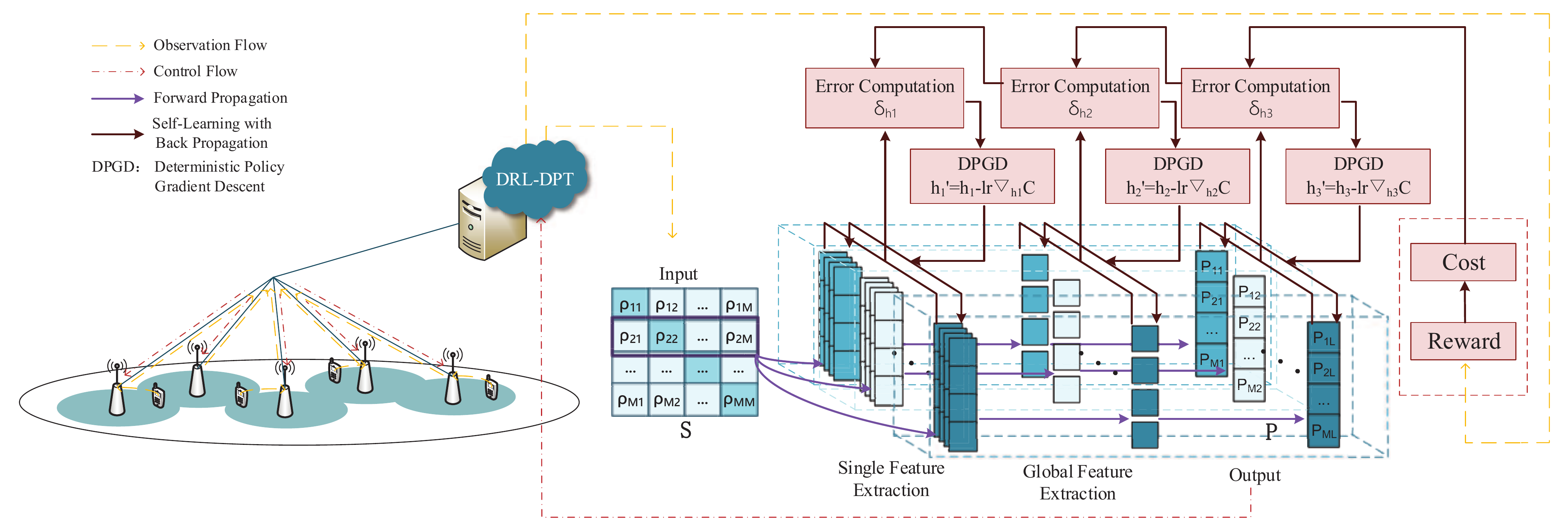}
\caption{The working architecture of DRL-DPT. DRL-DPT working on the MEC server mainly includes two processes: forward propagation through DNN to get continuous control and back propagation to adjust weights by deterministic policy gradient descent.}
\label{arch}
\end{figure*}

The structure of the proposed DRL-DPT is shown in Fig.~\ref{arch}.
It can be divided into $L$ drawers with same architecture transversely. Every drawer corresponds to a subframe while all of them share the same input.
This parted design is inspired by the fairness between SBSs and could reduce the complexity of DRL-DPT.
There are four layers in DRL-DPT, including input layer, single feature extracting layer, global feature extracting layer and output layer.
Their forward propagation functions are described in the following paragraphs.

\begin{enumerate}[(1)]

\item \textbf{\itshape Input layer:} At the beginning of each frame, the SBSs collect reference signals from their associated users to obtain necessary information about the wireless system.
    To reduce the data transmitted to the MEC server and to reduce the size of DRL-DPT, some pre-operations are manipulated at SBSs.
    The results can be regarded as partial observations and have the following expressions.
    \begin{equation}
    \begin{aligned}
    S_m = &\{\sum_{n=1}^{N_m}R_n^{th}, \sum_{n=1}^{N_m}S_{1,n}, \sum_{n=1}^{N_m}S_{2,n},\\
    &..., \sum_{n=1}^{N_m}S_{m-1,n}, \sum_{n=1}^{N_m}S_{m+1,n},..., \sum_{n=1}^{N_m}S_{M,n} \}
    \end{aligned}
    \label{eq7}
    \end{equation}
    To simplify the presentation, we rewrite the expression as
    \begin{equation}
    S_m = \{\rho_{m,1}, \rho_{m,2},..., \rho_{m,k},..., \rho_{m,M}\}
    \label{eq8}
    \end{equation}
    where $\rho_{m,k} = \left\{
                            \begin{array}{lr}
                            \sum_{n=1}^{N_m}R_n^{th}, k\!=\!m\\
                            \sum_{n=1}^{N_m}S_{k,n}, k\!\not=\! m
                            \end{array}
                        \right.$.
    $S_m$ contains the information about the required load and the suffered interferences of SBS $m$.

    The server receives these partial observations and integrates them as the final state input to DRL-DPT.
    The integrated state is expressed as
    \begin{equation}
    \textbf{S} = \begin{bmatrix}
                    \rho_{1,1} & \rho_{1,2} & \rho_{1,3} & ... & \rho_{1,M}\\
                    \rho_{2,1} & \rho_{2,2} & \rho_{2,3} & ... & \rho_{2,M}\\
                    ... & ... & ... & ... & ...\\
                    \rho_{M,1} & \rho_{M,2} & \rho_{M,3} & ... & \rho_{M,M}\\
                \end{bmatrix}
    \label{eq9}
    \end{equation}

    The input state is organized in the form of a $M \!\times\! M$ matrix and its diagonal elements represent the load of each SBS while its non-diagonal elements represent interferences from other SBSs.

\item \textbf{\itshape Single feature extracting layer:} $F \!\times\! L$ filters with size of $1 \!\times\! M$ are set up to extract the feature about the relative strength between different interferences as well as their influence to the specific SBS's load.
    $F$ is the number of filters set in one drawer.
    The filters in this layer has the same size as the mentioned partial states, so they can abstract input on SBS level and obtain a general impression.
    The inputs of the $M \!\times\! F \!\times\! L$ nodes in this layer are:
    \begin{equation}
    H_{1i}(m,f,l) = \sum_{k=1}^M h_1(k,f,l) \textbf{S}(m,k)
    \label{eq10}
    \end{equation}
    $h_1(k,f,l)$ is the $k$th weight of the $f$th filter in the $l$th drawer in this layer.
    And the outputs are
    \begin{equation}
    \centering
    \begin{aligned}
    H_{1o}(m,f,l) = act[H_{1i}(m,f,l)]\\
    1 \!\leq\! m \!\leq\! M, 1 \!\leq\! f \!\leq\! F, 1 \!\leq\! l \!\leq\! L
    \end{aligned}
    \label{eq11}
    \end{equation}
    where $act(\cdot)=sigmoid(\cdot)$ is the activation function we choose.

\item \textbf{\itshape Global feature extracting layer:} There are $F \!\times\! L$ filters with size of $M \!\times\! 1$ in this layer.
    They do comparisons between the extracted features of different SBSs to get details so that DRL-DPT can schedule the system wisely.
    The $M \!\times\! L$ nodes' inputs in this layer are expressed as
    \begin{equation}
    H_{2i}(m,l) = \sum_{f=1}^F h_2(m,f,l) H_{1o}(m,f,l)
    \label{eq12}
    \end{equation}
    where $h_2(m,f,l)$ is the $m$th weight of the $f$th filter in the $l$th drawer in the global feature extracting layer.
    Their outputs are
    \begin{equation}
    H_{2o}(m,l) = act[H_{2i}(m,l)]
    \label{eq13}
    \end{equation}

\item \textbf{\itshape Output layer:} Partial full-connections are operated to get the final output actions in this layer.
    The output layer has $M \!\times\! M \!\times\! L$ weights and $M \!\times\! L$ nodes in total.
    There are similar inputs and outputs in this layer as the former two:
    \begin{equation}
    H_{3i}(m,l) = \sum_{k=1}^M h_3(k,m,l) H_{2o}(m,l)
    \label{eq14}
    \end{equation}
    \begin{equation}
    H_{3o}(m,l) = act[H_{3i}(m,l)]
    \label{eq15}
    \end{equation}
    $h_3(k,m,l)$ is the $(k\!\times\!m)$th weight in the $l$th drawer.
    Actually, the output of DRL-DPT is the precise power allocation scheme, which can be expressed as
    \begin{equation}
    \textbf{P} = \begin{bmatrix}
                 P_{1,1} & P_{1,2} & P_{1,3} & ... & P_{1,L}\\
                 P_{2,1} & P_{2,2} & P_{2,3} & ... & P_{2,L}\\
                 ... & ... & ... & ... & ...\\
                 P_{M,1} & P_{M,2} & P_{M,3} & ... & P_{M,L}\\
                 \end{bmatrix}
    \label{eq16}
    \end{equation}
    where $P_{m,l} \!=\! P \!\cdot\! H_{3o}(m,l)$.

\end{enumerate}

Finally, DRL-DPT gets a $M \!\times\! L$ matrix which represent the learned optimal power allocations.
The MEC server informs SBSs with the corresponding elements in the output power matrix to realize ICIC by power control.

\subsection{Learning Mechanism}

We have described the forward propagation process of the actor above.
In this section, we explain the learning mechanism of DRL-DPT and give theoretical supports.
The key point is the design of the reward function.
In typical RL algorithms, the reward function is usually regarded as unknown, making the learning process slow and uncertain. However, in many scenarios, such as the considered wireless system, a lot of mature analyses and effective works can be found. It is meaningful to seek guidance from them. In this paper, we develop a deterministic target based on the common knowledge in wireless networks and give the theoretical presentation of reward function. Then a deterministic policy gradient descent algorithm is employed to adjust the parameterized policy, aiming at maximizing the reward. The dedicated reward function is consistent with the objective in equation (10). It saves the agent from conjecturing the reward, allowing more direct and quick learning.
The detailed process is illustrated as follow:
\begin{enumerate}[(1)]
\item \textbf{\itshape Reward function:} With the target of maximizing the system energy efficiency, we denote the reward function theoretically on the bases mentioned in section 3. The expression is
    \begin{equation}
    r = \gamma exp(\eta)
    \end{equation}
    where $\gamma$ is a positive constant coefficient. For specific state $\textbf{S}$ and action $\textbf{P}$, $r$ could be calculated.

    This design provides a certain representation for the rewards and it is related to the output actions. It determines the learning target clearly for the actor as well as indicating the learning approach. By using a policy gradient descent algorithm, the actor tunes its parameters iteratively to get higher rewards. The maximum could be obtained after enough iterations, meaning that the system is optimized in term of energy efficiency. This well-defined reward function essentially simplify the learning process by saving the agent from uncertain predictions. It provides a deterministic target which is in line with the optimization objective. It also shows the rules of action selecting so that the actor may know what to do more explicitly.

\item \textbf{\itshape Policy gradient descent:} To train the actor under the guidance of the interactive reward function means to maximize the reward $r$. In practice, we minimize the cost:
    \begin{equation}
    C = -r
    \end{equation}
    A deterministic policy gradient descent algorithm with error back propagation \cite{b19} is adopted.
    When a reward is fed back to DRL-DPT and converted to cost, the error of each layer is computed by

    \begin{subequations}
    \footnotesize
    \begin{gather}
    \delta_{h3}(m,l)\!=\! \frac{\partial C}{\partial H_{3o}(m,l)} \frac{\mathrm{d}act}{\mathrm{d}h_{3i}(m,l)}\\
    \delta_{h2}(m,l) \!=\! \sum_{k=1}^M [\delta_{h3}(k,l) h_3(m,k,l)]\frac{\mathrm{d}act}{\mathrm{d}H_{2i}(m,l)}\\
    \delta_{h1}(m,f,l) \!=\! \delta_{h2}(m,l) h_2(m,1,f,l) \frac{\mathrm{d}act}{\mathrm{d}H_{1i}(m,f,l)}
    \end{gather}
    \label{eq17}
    \end{subequations}

    Then the policy gradients of every weights in the actor network can be computed by
    \begin{subequations}
    \begin{gather}
    \nabla_{h_3(m_1,m_2,l)}C = \delta_{h3}(m_2,l) H_{2o}(m_1,l)\\
    \nabla_{h_2(m,1,f,l)}C = \delta_{h2}(m,l) H_{1o}(m,f,l)\\
    \nabla_{h_1(1,m,f,l)}C = \sum_{k=1}^M [\delta_{h1}(k,f,l) I(k,m)]
    \end{gather}
    \label{eq18}
    \end{subequations}

    Assuming the learning rate as $lr$, the weights are adjusted by the following rules:
    \begin{subequations}
    \begin{align}
    h_3' = h_3 - lr \nabla_{h_3}C\\
    h_2' = h_2 - lr \nabla_{h_2}C\\
    h_1' = h_1 - lr \nabla_{h_1}C
    \end{align}
    \label{eq19}
    \end{subequations}

    In this way, the actor performs a self-learning towards the target rule determinately.

\end{enumerate}

The entire procedure of using DRL-DPT to implement ICIC in wireless networks is concluded as algorithm 1.
\begin{algorithm}[h]
\caption{DRL-DPT}
\KwIn{$\textbf{S} = [S_1^T,S_2^T,...,S_M^T]^T$}
\KwOut{$\textbf{P} = [P_{m,t}]_{M \times L}(m \in \mathbb{M}, t \in \mathbb{L})$}
Initialize $C$ to $+\infty$\;
\Repeat{$C \le \epsilon$}{
\SetKw{Kwobs}{observe procedure}
\Kwobs{\:}\\
The MEC server receives $S_m$ from SBSs and orchestrates them to the input of DNN: $S_m \Rightarrow \textbf{S}$\;
\SetKw{Kwstr}{action learning procedure}
\Kwstr{\:}\\
Run DRL-DPT forward through (~\ref{eq10}) $\sim$ (~\ref{eq16})\;
Get the power allocation scheme $\textbf{P}$\;
\SetKw{Kwexe}{execute procedure}
\Kwexe{\:}\\
Set SBSs' transmit power according to $\textbf{P}$\;
\SetKw{Kwtra}{train procedure}
\Kwtra{\:}\\
evaluate the wireless network by\: $r = \gamma_1 exp(\eta)$\;
convert $r$ to cost: $C = -r$\;
compute the gradients by (~\ref{eq18})\;
adjust DNN by (~\ref{eq19})\;
}
\end{algorithm}
$\epsilon$ is a predefined threshold indicated the termination of the algorithm.

\section{Experiments}

\begin{figure}
\centering
\subfigure[Energy-efficiency of wireless networks of different users.]{
\begin{minipage}[b]{0.45\textwidth}
\centering
\includegraphics[width=7cm, height=6cm]{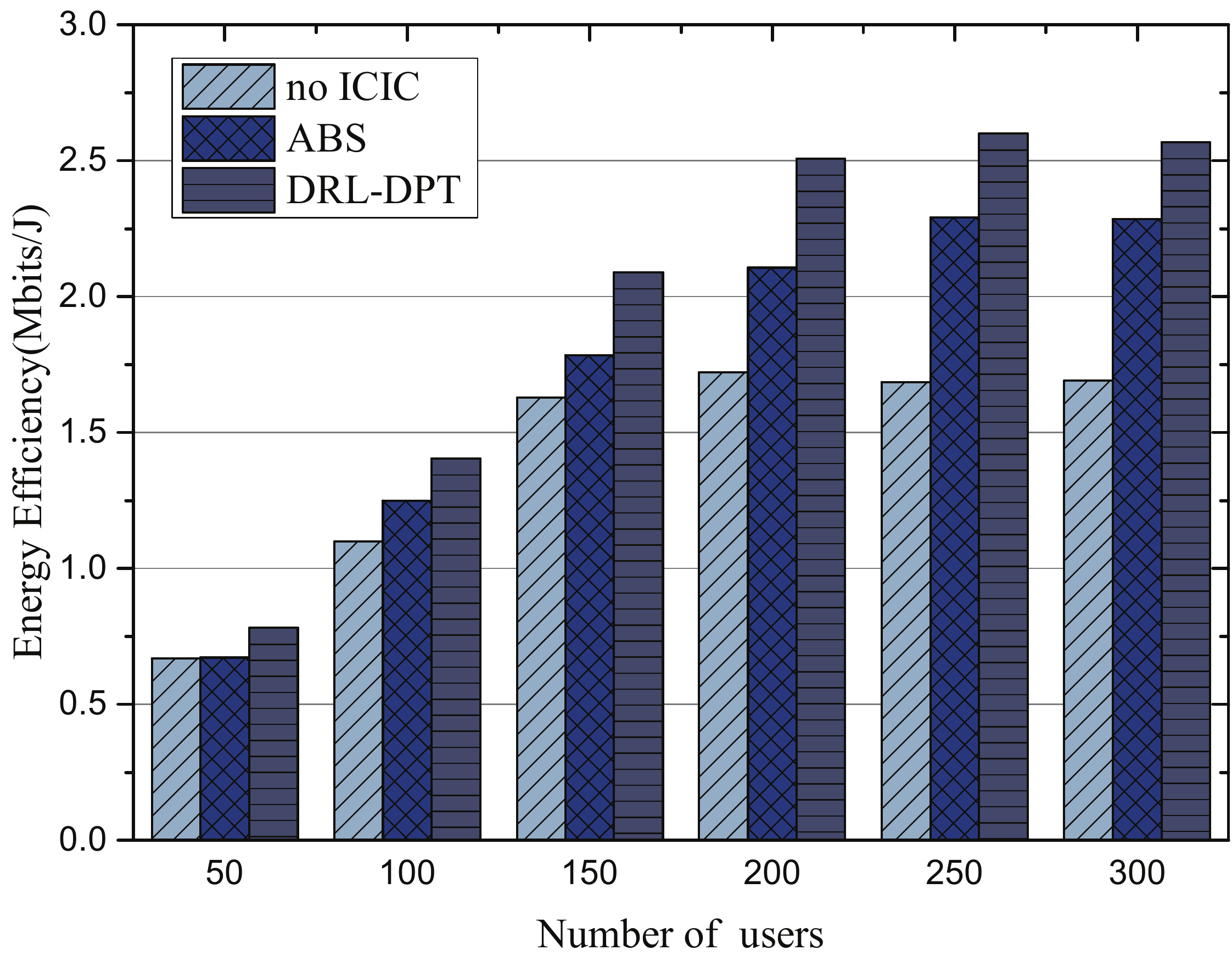}
\label{exp11}
\end{minipage}
}
\subfigure[Throughput of wireless networks of different users.]{
\begin{minipage}[b]{0.45\textwidth}
\centering
\includegraphics[width=7cm, height=6cm]{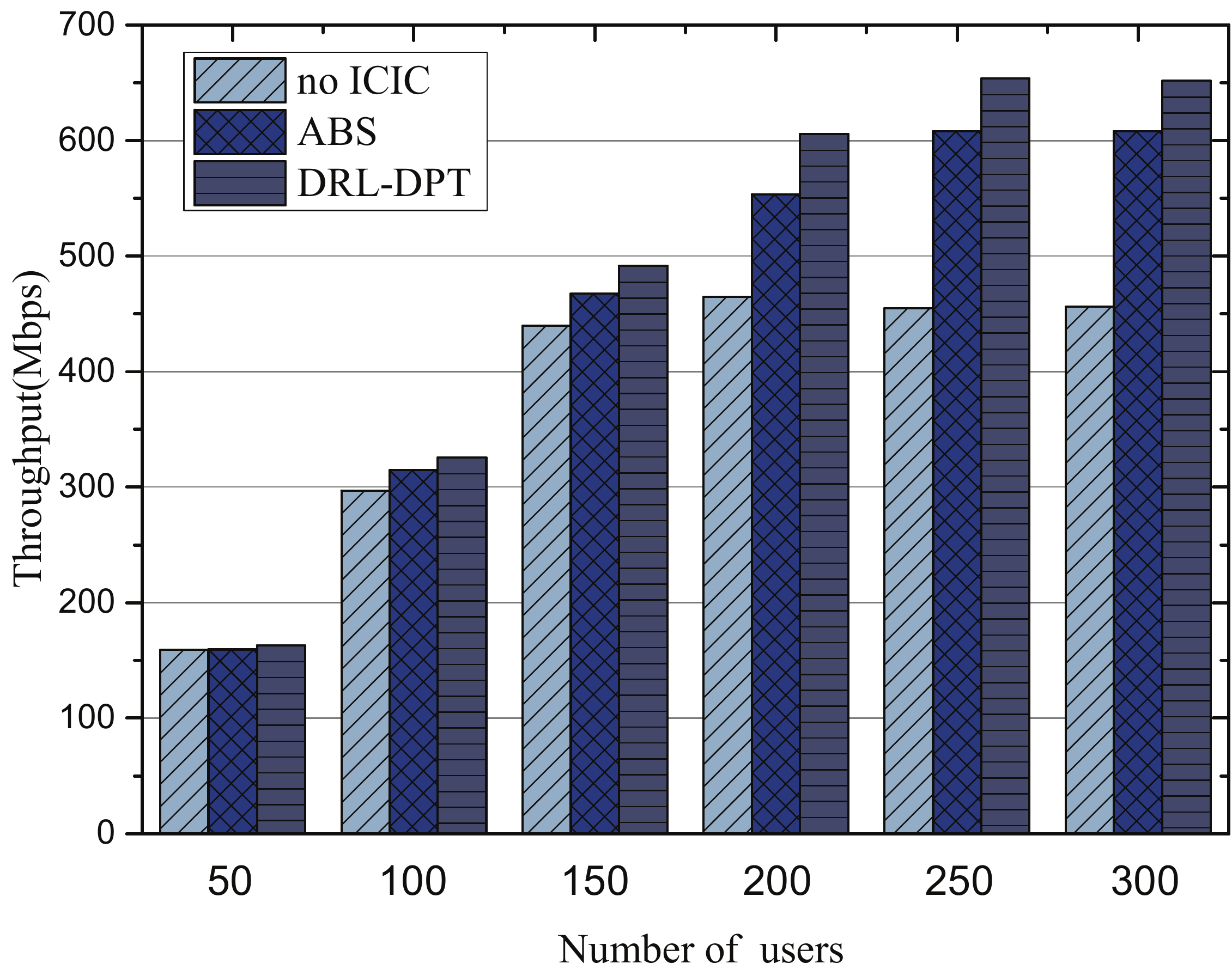}
\label{exp12}
\end{minipage}
}
\caption{The system performance of of wireless networks with different number of users using different ICIC schemes. SBSs are deployed according to a $5 \!\times\! 5$ grid model.}
\label{exp1}
\end{figure}

Experiments are performed to checkout the proposed DRL-DPT algorithm in terms of effectiveness and maneuverability. Some experimental parameters are given in Tab.~\ref{tab1}.
\begin{table}[h]
\centering
\caption{Experimental parameters.}
\begin{tabular}{|c|c|}
\hline
Traffic Model & Full Buffer Model\\
\hline
Transmission Power & 30dbm(SBSs)\\
\hline
System Bandwidth & 10MHz\\
\hline
SBS-user Path Loss Model & $140.7\!+\!26.7log_{10}(d)$\\
\hline
Cell Layer & $2 \!\times\! 2$ or $5 \!\times\! 5$ Grid Model\\
\hline
Distance Between SBSs & 50m\\
\hline
Learning Rate & 0.1\\
\hline
$\gamma$ & 0.5\\
\hline
\end{tabular}
\label{tab1}
\end{table}

In Fig.~\ref{exp1}, the system performance under different ICIC schemes are displayed. 25 SBSs are deployed according to the grid model and the number of users varies. We choose a traditional time-domain ICIC scheme based on ABS \cite{b20} and a power control scheme without ICIC as two baselines. As expected, the power control scheme without ICIC shows the worst system performance in terms of energy efficiency and throughput. It indicates that ICIC is necessary for wireless networks with densely deployed SBSs. Fig.~\ref{exp1} also shows that the proposed learning scheme outperforms the traditional ABS based schemes up to 15\% in terms of energy efficiency. As the number of users increases from 50 to 200, more performance gains are obtained with DRL-DPT. However, in simulation scenarios with 250 and 300 users, the system performance almost remains constant for all schemes. The reason is that the system runs into resource saturation, which means that there is no power available for power control operation.
\iffalse
Fig.~\ref{exp11} shows that the proposed learning method improves the networks' energy-efficiency significantly, especially when the users number growing and the interference getting heavier. While the networks without ICIC reach the bottleneck in terms of energy-efficiency, DRL-DPT provides networks the ability of keeping growing by better interference coordination. In fig.~\ref{exp12}, we compare the system throughput of different networks. It can be seen that, though we focus on the objective of energy-efficiency, we increase the throughput greatly as well. When there are more than 150 users in the network, the system offers no more capacity if no ICIC scheme is performed. ABS can improve this situation. And the proposed DRL-DPT could even achieve 10\% improvement more.
\fi

\begin{figure}
\centering
\subfigure[Energy-efficiency varies with iterations in 20-users network.]{
\begin{minipage}[b]{0.45\textwidth}
\centering
\includegraphics[width=8cm, height=6cm]{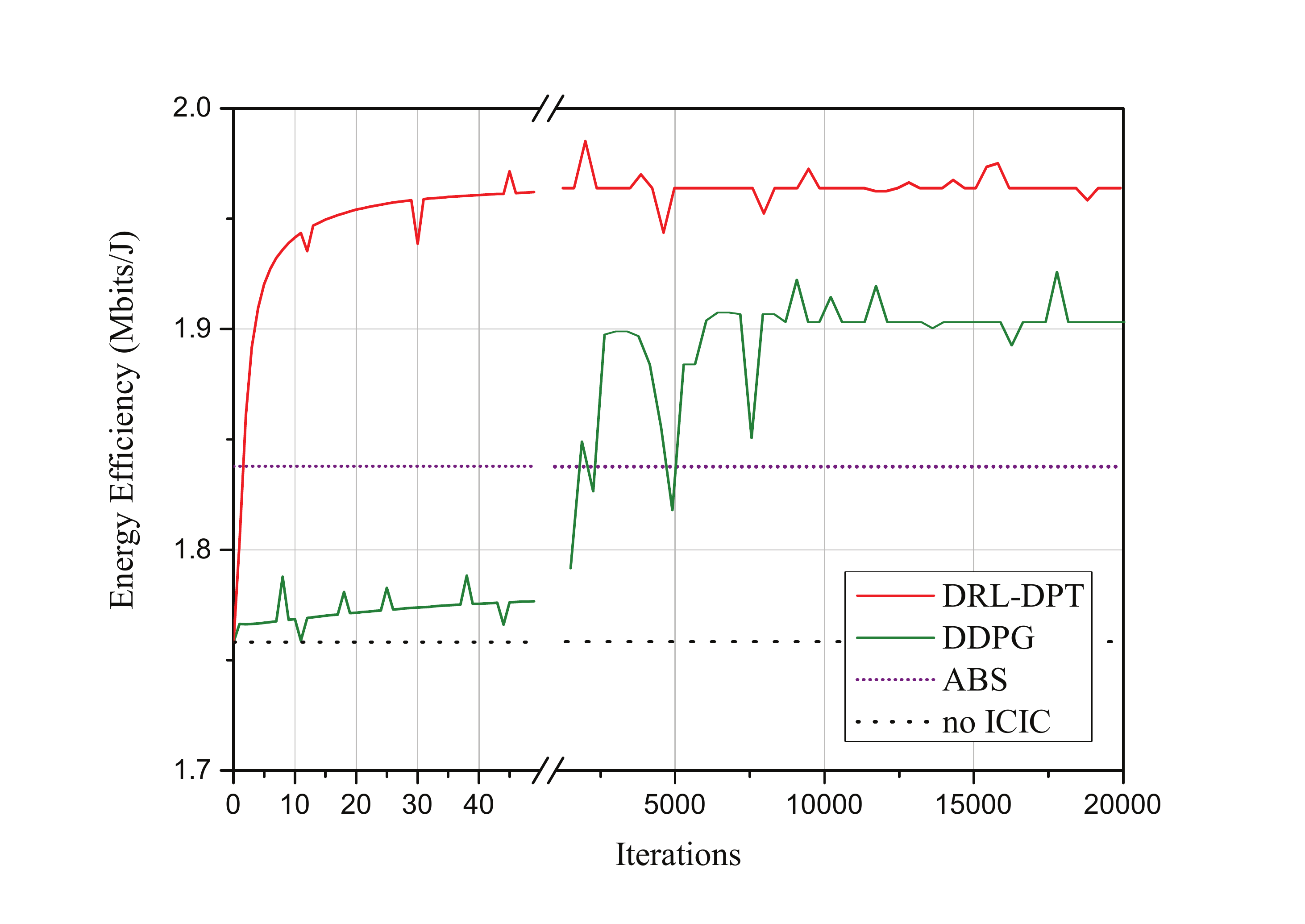}
\label{conv20}
\end{minipage}
}
\subfigure[Energy-efficiency varies with iterations in 100-users network.]{
\begin{minipage}[b]{0.45\textwidth}
\centering
\includegraphics[width=8cm, height=6cm]{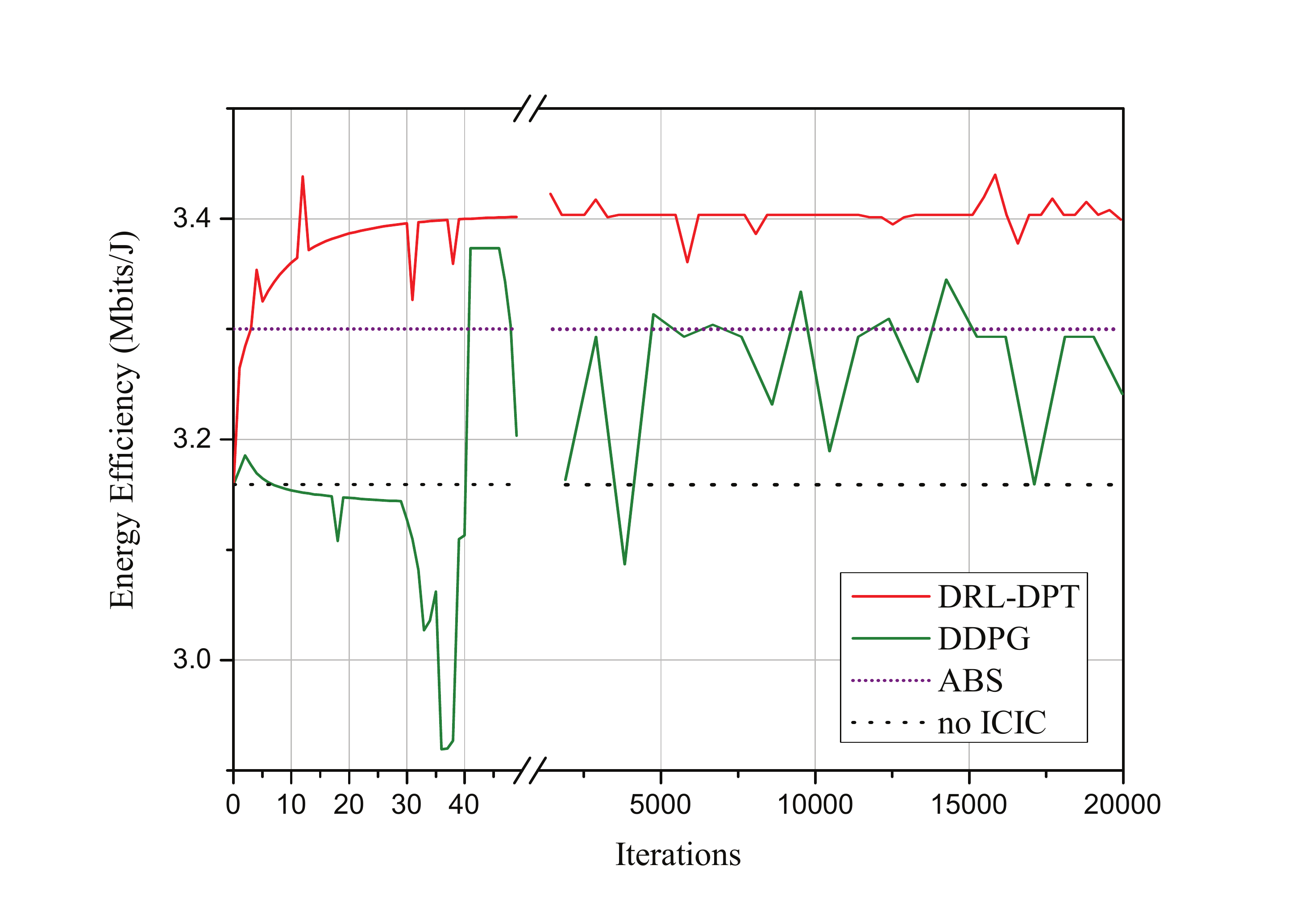}
\label{conv100}
\end{minipage}
}
\caption{The curves of energy efficiency of wireless networks with different ICIC schemes. It shows the convergence of DRL-DPT and DDPG as well as their effect on the system performance, with two baselines, i.e. ABS and no ICIC. The experiment is performed in a $2 \!\times\! 2$ cell grid model. ~\ref{conv20} considers 20 users in the network and ~\ref{conv100} considers 100 users. The figures are drawn with a break to show the convergence clearly.}
\label{conv}
\end{figure}

In order to show the convergence of DRL-DPT, a simple DDPG framework with a three-layer critic is also carried out to compare with. The critic predicts the system reward and guides the actor when selecting actions. Meanwhile, the critic also trains itself according to the actual feedback rewards to realize accurate predictions. Fig.~\ref{conv} shows the convergence of DRL-DPT and DDPG under different system scales. Fig.~\ref{conv20} shows the result when there are 20 users connected to the network. DRL-DPT improves the energy-efficiency quickly and gets stable in about 20 iterations, while DDPG learns slowly. When there are more users in the system, i.e. 100 users considered in fig.~\ref{conv100}, DRL-DPT can still converge within tens of iterations and stay stable. And under this condition, DDPG becomes more unstable and suffers from violent oscillation. These results reveal the significance of deterministic targets. DRL-DPT learns with the guidance of the deterministic target rules so that it behaves with a better sense of direction. To the contrary, DDPG has to predict the rules at first, so it takes much more time to find the optimal actions and suffers instability. With the pre-defined target rules, DRL-DPT can find the rewarding behavior quickly and accurately.

\iffalse
\begin{figure}[t]
\centering
\includegraphics[width=7cm, height=5.5cm]{conv20.png}
\caption{The curve of energy efficiency of wireless networks with different ICIC algorithms. It shows the convergence of DRL-DPT and DDPG as well as their effect on the system performance, with two baselines, i.e. ABS and no ICIC. The experiment is performed in a $2 \!\times\! 2$ cell grid model and 50 users are considered. The figure is drawn with a break to show the convergence clearly.}
\label{exp2}
\end{figure}
\fi

\section{Conclusion}

In this paper, we propose a novel deep reinforcement learning algorithm to implement the inter-cell interference coordination in DSCNs.
The proposed algorithm obtains good grades working in continuous domain with deterministic policy and target rule.
Some mature acknowledgements in wireless networks are adapted to DRL to implement a theoretical guidance.
Though the work is still primary and superficial, it is believed that RL technologies will bring huge energy to the future wireless networks.
And it is promising that our research could inspire more and more excellent works.

As future work, we will improve the framework to better adapt to the dynamics in wireless networks.
It is also valuable to find ways to reduce the system burden further.

\section*{ACKNOWLEDGMENT}
This work was supported in part by the National Science Foundation of China (No.91538203, 61390513, 61771445) and the Fundamental Research Funds for the Central Universities.


\begin{thebibliography}{00}
\bibitem{b1} Cisco Visual Networking Index: Global Mobile Data Traffic Forecast Update, 2015-2020 White Paper, document 1454457600805266, Cisco, San Jose, CA, USA, Feb. 2016. [Online]. Available: \url{https://www.cisco.com/c/dam/m/en_in/innovation/enterprise/assets/mobile-white-paper-c11-520862.pdf}
\bibitem{b2} Bhushan N, Li J, Malladi D, et al. Network densification: the dominant theme for wireless evolution into 5G[J]. IEEE Communications Magazine, 2014, 52(2): 82-89.
\bibitem{b3} Sutton R S, Barto A G. Reinforcement learning: An introduction[M]. Cambridge: MIT press, 1998.
\bibitem{b4} Mnih V, Kavukcuoglu K, Silver D, et al. Human-level control through deep reinforcement learning[J]. Nature, 2015, 518(7540): 529.
\bibitem{b5} Lillicrap T P, Hunt J J, Pritzel A, et al. Continuous control with deep reinforcement learning[J]. arXiv preprint arXiv:1509.02971, 2015.
\bibitem{b6} Auer G, Giannini V, Desset C, et al. How much energy is needed to run a wireless network?[J]. IEEE Wireless Communications, 2011, 18(5).
\bibitem{b7} Silver D, Schrittwieser J, Simonyan K, et al. Mastering the game of go without human knowledge[J]. Nature, 2017, 550(7676): 354.
\bibitem{b8} Kim H, de Veciana G, Yang X, et al. Alpha-optimal user association and cell load balancing in wireless networks[C]//INFOCOM, 2010 Proceedings IEEE. IEEE, 2010: 1-5.
\bibitem{b9} Jiming C, Peng W, Jie Z. Adaptive soft frequency reuse scheme for in-building dense femtocell networks[J]. China Communications, 2013, 10(1): 44-55.
\bibitem{b10} Son K, Kim H, Yi Y, et al. Base station operation and user association mechanisms for energy-delay tradeoffs in green cellular networks[J]. IEEE journal on selected areas in communications, 2011, 29(8): 1525-1536.
\bibitem{b11} Ashraf I, Boccardi F, Ho L. Sleep mode techniques for small cell deployments[J]. IEEE Communications Magazine, 2011, 49(8).
\bibitem{b12} Buda T S, Assem H, Xu L, et al. Can machine learning aid in delivering new use cases and scenarios in 5G?[C]//Network Operations and Management Symposium (NOMS), 2016 IEEE/IFIP. IEEE, 2016: 1279-1284.
\bibitem{b13} Watkins C J C H, Dayan P. Q-learning[J]. Machine learning, 1992, 8(3-4): 279-292.
\bibitem{b14} Kansal A, Zhao F. Fine-grained energy profiling for power-aware application design[J]. ACM SIGMETRICS Performance Evaluation Review, 2008, 36(2): 26-31.
\bibitem{b15} Zhan Y, Ammar H B. Theoretically-grounded policy advice from multiple teachers in reinforcement learning settings with applications to negative transfer[J]. arXiv preprint arXiv:1604.03986, 2016.
\bibitem{b16} Nguyen V M, Kountouris M. Performance limits of network densification[J]. IEEE Journal on Selected Areas in Communications, 2017, 35(6): 1294-1308.
\bibitem{b17} Hu Y C, Patel M, Sabella D, et al. Mobile edge computing-A key technology towards 5G[J]. ETSI white paper, 2015, 11(11): 1-16.
\bibitem{b18} Zappone A, Sanguinetti L, Bacci G, et al. Energy-efficient power control: A look at 5G wireless technologies[J]. IEEE Transactions on Signal Processing, 2016, 64(7): 1668-1683.
\bibitem{b19} Chauvin Y, Rumelhart D E. Backpropagation: theory, architectures, and applications[M]. Psychology Press, 2013.
\bibitem{b20} Lopez-Perez D, Guvenc I, De la Roche G, et al. Enhanced intercell interference coordination challenges in heterogeneous networks[J]. IEEE Wireless Communications, 2011, 18(3).
\end{thebibliography}
\end{document}